\documentclass[english,twocolumn]{revtex4}
\usepackage[T1]{fontenc}
\usepackage[latin9]{inputenc}
\usepackage{bm}
\usepackage{amstext}
\usepackage{amssymb}
\usepackage{graphicx}
\usepackage{esint}

\makeatletter

\providecommand{\tabularnewline}{\\}

\@ifundefined{textcolor}{}
{%
 \definecolor{BLACK}{gray}{0}
 \definecolor{WHITE}{gray}{1}
 \definecolor{RED}{rgb}{1,0,0}
 \definecolor{GREEN}{rgb}{0,1,0}
 \definecolor{BLUE}{rgb}{0,0,1}
 \definecolor{CYAN}{cmyk}{1,0,0,0}
 \definecolor{MAGENTA}{cmyk}{0,1,0,0}
 \definecolor{YELLOW}{cmyk}{0,0,1,0}
}

\makeatother

\usepackage{babel}
\begin{document}

\title{Origin of Long Lived Coherences in Light-Harvesting Complexes}

\author{Niklas Christensson$^{1}$, Harald F. Kauffmann$^{1,2}$, T\~{o}nu Pullerits$^{3}$,
and Tom\'{a}\v{s} Man\v{c}al$^{4}$ }

\affiliation{$^{1}$Faculty of Physics, University of Vienna, Strudlhofgasse 4,
1090 Vienna, Austria}

\affiliation{$^{2}$Faculty of Physics, Vienna University of Technology, 1040
Vienna, Austria}

\affiliation{$^{3}$Department of Chemical Physics, Lund University, P.O. Box
124, SE-22100 Lund, Sweden }

\affiliation{$^{4}$Institute of Physics, Faculty of Mathematics and Physics,
Charles University, Ke Karlovu 5, Prague 121 16, Czech Republic }
\begin{abstract}
A vibronic exciton model is developed to investigate the origin of
long lived coherences in light-harvesting complexes. Using experimentally
determined parameters and uncorrelated site energy fluctuations, the
model predicts oscillations in the nonlinear spectra of the Fenna-Matthews-Olson
(FMO) complex with a dephasing time of 1.3 ps at 77 K. These oscillations
correspond to the coherent superposition of vibronic exciton states
with dominant contributions from vibrational excitations on the same
pigment. Purely electronic coherences are found to decay on a 200
fs timescale. 
\end{abstract}
\maketitle
\begin{figure}
\includegraphics[clip,width=1\columnwidth]{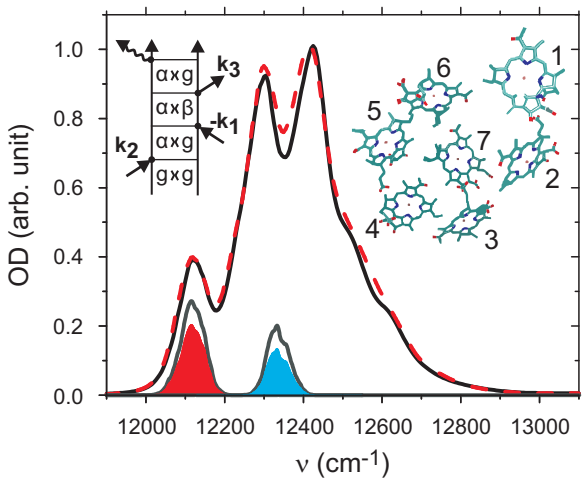}

\caption{\label{fig:Fig1}Linear absorption spectrum at $77$ K for the exciton
model (black) and the vibronic exciton model (red dash). The gray
lines show the distribution of renormalized transition frequencies
weighted by the transition strength (scaled by $2/3$) for state $1$
and $4$ in the vibronic exciton model. The filled areas illustrate
the relative contribution from electronic (red) and vibronic (blue)
excitations on pigment $3$. The spectra of the vibronic exciton model
have been shifted by the reorganization energy of the vibrational
mode ($-9.25$ cm$^{-1}$) for comparison. The insets show the Feynman
diagram illustrating the non-rephasing excited state coherence pathway
in Eq. (12), and the arrangement of the $7$ BChls in FMO (\emph{C.
tepidum}) \cite{Tronrud}. This figure was generated using the VMD
software \cite{Humphrey1996a}. }
\end{figure}

The role of quantum mechanics in biological processes, as well as
photosynthetic light-harvesting, has been of interest for a long time
\cite{Davydov,PhotosynthExcitons,Kuehn2002a}. The topic received
renewed attention after the observation of long lived oscillations
in the two-dimensional (2D) spectra of the Fenna-Matthews-Olson (FMO)
protein pigment complex \cite{Engel2007a}. These oscillations were
interpreted as a signature of electronic coherences between the delocalized
energy eigenstates of the complex, and it was argued that their slow
dephasing could enhance the efficiency of energy transfer between
the chlorosome antenna and the reaction center \cite{Engel2007a,Ishizaki2009a}.
Subsequent studies revealed that the oscillations in FMO have a dephasing
time of $1.2$ ps at $77$ K \cite{Hayes2011a}, and that such oscillations
are a common feature in light-harvesting complexes \cite{Calhoun2009a,Collini2009b}.
Based on the known structure of FMO (Fig. \ref{fig:Fig1}), simulations
employing formally exact equations of motions for the reduced density
operator found oscillations with dephasing times of $200-300$ fs
- clearly shorter than the experimental observation \cite{Ishizaki2009a,Chen2011a,Hein2011,Nalbach2011a}.
Significantly longer dephasing times were found if the transition
frequency fluctuations (static or dynamic) of the different pigments
are assumed to be correlated \cite{Abramavicius2011a,Caycedo2011b}.
However, theoretical studies using molecular dynamics simulations
of the interaction between electronic and nuclear degrees of freedom
(DOF) have not been able to confirm the existence of such correlations
\cite{Olbrich2011a}, or long lived electronic coherences \cite{Shim2011a}.
To what extent the long lived oscillations in the experiments reflect
electronic coherences, or if they influence the transport of energy
across the complex \cite{Engel2007a,Panitchayangkoon2011a}, thus
remains an open question. 

Analysis of excitation dynamics in molecular aggregates typically
employs a reduced description, where the electronic DOF and their
mutual couplings are treated explicitly, and the nuclear modes of
the pigments and protein are treated as a heat bath \cite{MayKuehnBook,Adolphs2006a}.
The presence of underdamped vibrational modes in the bath produces
oscillatory signatures in 2D spectra which are similar to the modulations
predicted for electronic coherences \cite{Christensson2011a}. The
dephasing time of such nuclear coherence are of the order of several
picoseconds, and they can often be treated as completely undamped
on the time scale of a typical 2D experiment. Low temperature fluorescence
line narrowing (FLN) experiments on FMO have revealed a large number
of vibrational modes in the range of $0-350$ cm$^{-1}$\cite{Wendling2000a}.
However, the oscillations seen in the 2D spectra of FMO cannot be
directly related to simple nuclear wavepackets, because the frequency
of the oscillations does not match any of the vibrational frequencies,
and the Huang-Rhys factors of the modes are too low. On the other
hand, recent simulations have revealed unexpected effects on the electronic
structure and dynamics if vibrational modes are explicitly included
in the system \cite{Polyutov2011a}. Motivated by these results, we
develop a vibronic exciton Hamiltonian in so called one particle approximation
\cite{Philpott1971a,Spano2002a} for FMO, in which one vibrational
mode on each monomer is treated explicitly. We will show below that
this model predicts oscillations in the 2D spectra of FMO with $1.3$
ps dephasing times at $77$ K, and that these long lived coherences
can be traced to superpositions of vibronic exciton states located
on the same pigment.

The total Hamiltonian of a molecular aggregate in contact with the
environment is partitioned in a standard way into system, bath and
system-bath interaction terms, $H=H_{S}+H_{B}+H_{SB}$. The system
Hamiltonian describes the $Q_{y}$ transition on each bacteriochlorophyll
(BChl) in FMO (see Fig. \ref{fig:Fig1}) with the vibrational progression
of a single vibrational mode. Including the resonance coupling between
the transitions, the system Hamiltonian reads
\begin{equation}
H_{S}=\sum_{n,\nu,m,\nu^{\prime}}[\delta_{nm}\delta_{\nu\nu^{\prime}}(E_{n}+\nu\hbar\omega_{0})+J_{n,\nu;m,\nu^{\prime}}]|n,\nu\rangle\langle m,\nu^{\prime}|,\label{eq:HS}
\end{equation}
where $E_{n}$ is the transition frequency of pigment $n$ (site energy),
$\omega_{0}$ is the vibrational frequency, and $\nu$ is the quantum
number of the vibrational mode. The coupling energy $J_{n,\nu;m,\nu^{\prime}}$
between the individual transitions can be expressed via the electronic
resonance coupling $J_{nm}$ \cite{Adolphs2006a} and the Franck-Condon
amplitudes of the vibrational mode \cite{Spano2002a} $J_{n,\nu;m,\nu^{\prime}}=\langle\nu|0\text{\ensuremath{\rangle\langle}}\nu'|0\rangle$.
The eigenvalues and wave-functions of $H_{S}$ are given by $\hbar\omega_{\alpha}$
and $|\alpha\rangle=\sum_{n,\nu}c_{n,\nu}^{\alpha}|n,\nu\rangle$,
respectively. The bath Hamiltonian, $H_{B}$, is described as a collection
of independent harmonic oscillators, for which the system-bath interaction
is given by $H_{SB}=\sum_{n,\nu}\omega d_{n}(\omega)\tilde{q}_{n}|n,\nu\rangle\langle n,\nu|.$
Here $\tilde{q}$ is a generalized coordinate of the environment,
and $d(\omega)$ is the displacement of the excited state relative
to the ground state. We assume that the system-bath Hamiltonian does
not depend on the state of the vibrational mode $(\nu)$, implying
that a vibrational coherence on an isolated monomer is undamped. Assuming
equal but uncorrelated system-bath interaction for the different pigments,
the energy gap correlation function in the local basis is given by
$C_{nm}(t)=\omega^{2}d_{n}d_{m}\langle\tilde{q}_{n}(t)\tilde{q}_{m}(0)\rangle=C_{0}(t)\delta_{nm}.$
When the interaction of the system with the environmental modes (i.e.
all except those treated explicitly in the system Hamiltonian) is
weak, it is advantageous to perform calculations in the eigenstate
basis of the system Hamiltonian. The correlation function of the energy
gap in the eigenstate representation can be expressed via the expansion
coefficients, $c_{n,\nu}^{\alpha}$, and the correlation function
of each excitation in the local basis $C_{0}(t)$,
\begin{equation}
\begin{array}{c}
C_{\alpha\beta}(t)=C_{0}(t)\Big\{\sum_{n,\nu}(c_{n,\nu}^{\alpha})^{2}(c_{n,\nu}^{\beta})^{2}\\
+\sum\limits _{n,\nu\neq\nu^{\prime}}[(c_{n,\nu}^{\alpha})^{2}(c_{n,\nu^{\prime}}^{\beta})^{2}+(c_{n,\nu^{\prime}}^{\alpha})^{2}(c_{n,\nu}^{\beta})^{2}]\Big\}=C_{0}(t)\gamma_{\alpha\beta}
\end{array}
\end{equation}
The dephasing dynamics of a coherent superposition between the eigenstates
$\alpha$ and $\beta$ is determined by the lineshape function $g_{\alpha\beta}(t)=\int_{0}^{t}{\rm d}\tau\int_{0}^{\tau}{\rm d}\tau^{\prime}\gamma_{\alpha\beta}C_{0}(\tau^{\prime})$.
The correlation function $C_{0}(t)$ is connected to the the spectral
density $\tilde{C}^{\prime\prime}(\omega)$ via a Fourier transform
(see e.g. Ref. \cite{MukamelBook}). In addition to dephasing, the
system-bath interaction leads to relaxation between the eigenstates
of the system. We use Redfield theory \cite{MayKuehnBook} (Markov
and secular approximation), where the relaxation rate is given by
\begin{equation}
\begin{array}{c}
k_{\alpha\rightarrow\beta}=2\pi\sum_{n,\nu,m,\nu^{\prime}}\delta_{nm}c_{n,\nu}^{\alpha}c_{m,\nu^{\prime}}^{\alpha}c_{n,\nu}^{\beta}c_{m,\nu^{\prime}}^{\beta}\\
\times\Big\{(1+n(\omega_{\alpha\beta}))\tilde{C}^{\prime\prime}(\omega_{\alpha\beta})+n(-\omega_{\alpha\beta})\tilde{C}^{\prime\prime}(-\omega_{\alpha\beta})\Big\},
\end{array}\label{eq:Cab}
\end{equation}
where $\tilde{C}^{\prime\prime}(\omega_{\alpha\beta})>0$ and $n(\omega_{\alpha\beta})$
is the Bose-Einstein distribution function. The total relaxation rate
from a level $\alpha$, $\Gamma_{\alpha}=\frac{1}{2}\sum_{\gamma\neq\alpha}k_{\alpha\rightarrow\gamma}$,
determines the lifetime broadening of this exciton state. Assuming
$k_{{\rm B}}T<\hbar\omega_{0}$, the linear absorption spectrum can
be calculated as \cite{Adolphs2006a} 
\begin{equation}
OD(\omega)\propto\omega\langle\sum_{\alpha}|\mu_{\alpha0}|^{2}Re\int\limits _{0}^{\infty}{\rm d}t\ e^{-g_{\alpha\alpha}(t)-\Gamma_{\alpha}t-i(\omega_{\alpha0}-\omega)t}\rangle_{\Delta,\Omega},
\end{equation}
where $\alpha$ runs over all exciton levels and the transition dipole
moments $\bm{\mu}_{\alpha0}$ are given by $\bm{\mu}_{\alpha0}=\sum_{n,\nu}c_{n,\nu}^{\alpha}\bm{\mu}_{n}\langle\nu|0\rangle.$
Here $\langle\dots\rangle_{\Delta,\Omega}$ denotes the average over
a random distribution of pigment energies and orientations of complexes. 

To simulate the oscillations in a third order experiment (i.e. 2D
spectra) we adopt the doorway window representation \cite{Zhang1998a}.
Of all Liouville pathways contributing to the signal, only those involving
a coherence between two levels in the excited state will give rise
to oscillations during the waiting time $t_{2}$. Without the loss
of generality, we focus on the non-rephasing coherence pathways illustrated
in Fig. \ref{fig:Fig1}, which give rise to oscillations along the
diagonal in the non-rephasing 2D spectrum \cite{Calhoun2009a}. The
response function for this pathway is given by 
\begin{equation}
R_{\alpha\beta,0}=\langle\langle(\bm{\mu}_{\alpha0})^{2}(\bm{\mu}_{\beta0})^{2}\rangle_{\Omega}G_{\alpha}(t_{3})G_{\alpha\beta}^{(2)}(t_{2})G_{\alpha}(t_{1})\rangle_{\Delta},\label{eq:Rab}
\end{equation}
where $G_{\alpha}(t)=e^{-i\omega_{\alpha0}t-\Gamma_{\alpha}t-g_{\alpha\alpha}(t)}$,
and $G_{\alpha\beta}^{(2)}(t_{2})=e^{-i\omega_{\alpha\beta}t_{2}}e^{-g_{\alpha\alpha}(t_{2})-g_{\beta\beta}(t_{2})+2g_{\alpha\beta}(t_{2})-(\Gamma_{\alpha}+\Gamma_{\beta})t_{2}}$.
In this work we use the site energies ($E_{n}$ ) and resonance couplings
($J_{nm}$) for FMO \emph{Chlorobium tepidum} from Ref. \cite{Adolphs2006a},
and the analytical formula for the overdamped part of the spectral
density, $\tilde{C}^{\prime\prime}(\omega)$, extracted from a FLN
experiment \cite{Wendling2000a}. The direction of the transition
dipole moments were taken from the protein data bank file 3ENI \cite{Tronrud}.
FLN experiments have identified $30$ vibrational modes in FMO, and
the strongest feature in the spectrum arises from three modes around
$185$ cm$^{-1}$ \cite{Wendling2000a}. To retain a simple description,
we treat this cluster of modes as one effective mode with a frequency
of $\omega_{0}=185$ cm$^{-1}$ and a Huang-Rhys factor of $0.05$.
For calculations with the exciton model, the vibrational mode was
included as a underdamped contribution in the spectral density. In
all calculations presented in this paper, we sampled the pigment transition
energies from a Gaussian distribution with a FWHM of $80$ cm$^{-1}$. 

\begin{figure}
\includegraphics[clip,width=1\columnwidth]{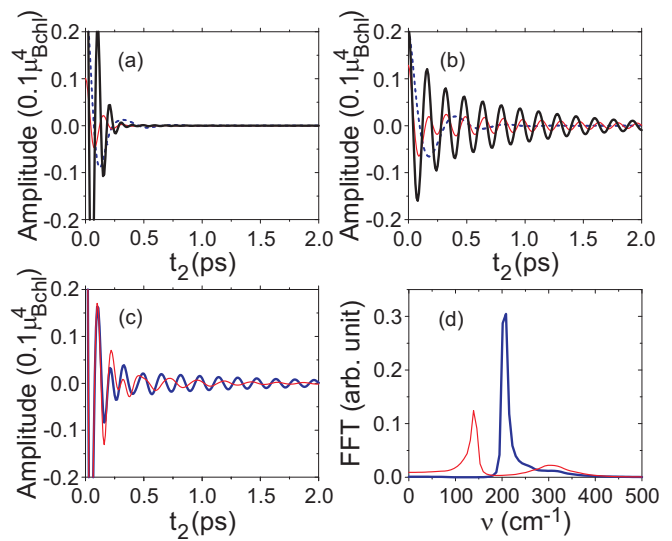}

\caption{\label{fig:Fig2}Amplitude of the real part of the non-rephasing coherence
pathways involving the lowest state. a) $Re\langle R_{1\beta}\rangle_{\Omega,\Delta}$
with $\beta=2$ (blue dash), $\beta=3$ (red thin solid), and $\beta=4$
(black solid) for the exciton model. b) $Re\langle R_{1\beta}\rangle_{\Omega,\Delta}$
with $\beta=2$ (blue dash), $\beta=3$ (red thin solid), and $\beta=4$
(black solid) for the vibronic exciton model. c) Sum of all non-rephasing
coherence pathways, $\sum_{\alpha\beta}Re\langle R_{\alpha\beta}\rangle_{\Omega,\Delta}$
, giving rise to signal in the range $12100\pm30$ cm$^{-1}$ for
the vibronic exciton model with $v_{0}=185$ cm$^{-1}$ (blue) and
$v_{0}$=117 cm$^{-1}$ (red) at $77$ K. The initial value of the
signal is $0.48$ and the first minimum at $t_{2}=0.04$ ps has an
amplitude of $-0.23$. The non-rephasing stimulated emission signal
calculated for the same parameters and spectral range has an initial
value of $0.17$. d) Power spectrum of the Fourier transform of the
signals in (c) starting from 0.2 ps. }
\end{figure}

The vibronic exciton and exciton model predict very similar linear
optical properties as illustrated by the simulated linear absorption
spectra shown in Fig. \ref{fig:Fig1}. Figure \ref{fig:Fig2}(a) shows
the time evolution of coherences involving the lowest state of the
exciton model. During $t_{2}$ the signals oscillate with frequencies
corresponding to the splitting between the exciton levels and are
completely damped after $400$ fs. The strong damping can be readily
understood from Eqs. (\ref{eq:Cab}) and (\ref{eq:Rab}). For the
exciton model we find that the cross-correlation term $g_{\alpha\beta}$
is small and the coherence pathways decay mainly with $\exp(-g_{\alpha\alpha}(t_{2})-g_{\beta\beta}(t_{2}))$.
If a (static or dynamic) correlation of the transition frequency fluctuations
in the site basis is assumed (ad hoc), the cross-correlation functions
in the exciton basis become larger and enable longer dephasing times.

The coherence pathways involving the lowest levels of the vibronic
exciton model are shown in Fig. \ref{fig:Fig2}(b). The vibronic exciton
coherences are remarkably long lived, and the signals show only minor
damping on a $2$ ps timescale. The long dephasing time of the coherences
in the vibronic exciton model can be understood by inspection of the
expansion factors $\langle|c_{n,\nu}^{\alpha}|^{2}\rangle_{\Delta}$
given in Tab. \ref{tab:Tab1} and the transition frequency distributions
shown in Fig. \ref{fig:Fig1}. For instance, state $1$ corresponds
to $75$ \% to an excitation of the $\nu=0$ transition of pigment
$3$, while state $4$ has a large contribution of vibrational excitation
($\nu=1$) on the same pigment. As discussed above, the system-bath
interaction is independent of the state of the vibrational mode, and
these two vibronic exciton levels will therefore experience highly
correlated fluctuations, resulting in slow dephasing of the $1,4$
coherence. Despite the large contribution from the vibrational excitation
to state $4$, it has a transition dipole moment which is comparable
to that of the other vibronic exciton levels. For non-interacting
pigments, only the zero-phonon state ($\nu=0$) has a significant
transition dipole moment. The strong transition dipole moment here
is the result of intensity borrowing from the electronic transitions
on the other pigments. As illustrated in Fig. \ref{fig:Fig1}, state
$4$ of the vibronic exciton model exhibits both properties needed
for the generation of long lived coherences: a large contribution
of vibrational excitation of pigment $3$, and a strong transition
dipole moment enabled by intensity borrowing. The combination of both
effects leads to long lived oscillations on the red edge of the linear
absorption spectrum as shown in Fig. \ref{fig:Fig2}(c). The oscillations
show a bi-phasic behavior, where the initial $~200$ fs decay of the
oscillation is due to the decay of coherences between vibronic exciton
states localized on different pigments (like the $1,2$ coherence,
see Fig. \ref{fig:Fig2}(b) and Tab. \ref{tab:Tab1}), while the long
lived oscillations reflect coherences between vibronic exciton states
localized on the same pigment. A fit to the oscillations give a dephasing
time of $1.3$ ps. By comparing to the stimulated emission signal
calculated for the same spectral range, we estimate the modulation
of the total signal, including stimulated emission and ground state
bleach, to be $5-10$ \% for $t_{2}>0.3$ ps. 

\begin{table}
\begin{tabular}{|c|c|c|c|c|}
\hline 
$\langle|c_{n,\nu}^{\alpha}|^{2}\rangle_{\Delta}$ & $\alpha=1$ & $\alpha=2$ & $\alpha=3$ & $\alpha=4$\tabularnewline
\hline 
\hline 
$n=1,\ \nu=0$ & $0.0$ & $0.03$ & $0.58$ & $0.15$\tabularnewline
\hline 
$n=1,\ \nu=1$ & $0.0$ & $0.0$ & $0.0$ & $0.0$\tabularnewline
\hline 
$n=3,\ \nu=0$ & $\bm{0.75}$ & $0.2$ & $0.03$ & $0.0$\tabularnewline
\hline 
$n=3,\ \nu=1$ & $0.0$ & $0.0$ & $0.2$ & $\bm{0.67}$\tabularnewline
\hline 
$n=4,\ \nu=0$ & $0.21$ & $0.51$ & $0.01$ & $0.01$\tabularnewline
\hline 
$n=4,\ \nu=1$ & $0.0$ & $0.0$ & $0.0$ & $0.01$\tabularnewline
\hline 
$\langle\mu_{\alpha}^{2}\rangle_{\Delta}(\mu_{BChl}^{2})$ & $\bm{0.87}$ & $0.58$ & $1.3$ & $\bm{0.57}$\tabularnewline
\hline 
$\langle E_{\alpha}\rangle_{\Delta}-\langle E_{1}\rangle_{\Delta}({\rm cm^{-1})}$ & 0.0 & $105$ & $175$ & $217$\tabularnewline
\hline 
\end{tabular}

\caption{\label{tab:Tab1}Contributions of selected basis excitations ($|c_{n,\nu}^{\alpha}|^{2}$)
to the four first vibronic exciton states averaged over energetic
disorder. The numbering of the pigments is defined in Fig. \ref{fig:Fig1}
and is the same as in Ref. \cite{Adolphs2006a}. The two bottom rows
show the averaged transition strength in units of the BChl monomer,
and the average energy differences between the vibronic exciton levels,
respectively. }
\end{table}

The Fourier transform of the signal in Fig. \ref{fig:Fig2}(c) is
shown in Fig. \ref{fig:Fig2}(d). The oscillation frequency of $205$
cm$^{-1}$ is higher than $\omega_{0}$ and also higher than the frequency
observed in the experiment ($160$ cm$^{-1}$) \cite{Hayes2011a}.
The oscillation frequency depends on transition energies, electronic
couplings and vibrational frequencies according to Eq. (\ref{eq:HS}).
Figs. \ref{fig:Fig2}(c) and (d) compares the oscillations on the
red edge of the spectrum for two different vibrational frequencies
found in the FLN experiment, $\omega_{0}=185$ cm$^{-1}$ and $\omega_{0}=117$
cm$^{-1}$. For $\omega_{0}=117$ cm$^{-1}$, the oscillations have
a frequency of $140$ cm$^{-1}$ and a shorter dephasing time as compared
to $\omega_{0}=185$ cm$^{-1}$. The dephasing time of the coherences
depend on the amount of vibrational character of the vibronic exciton
states, and detailed analysis of the oscillations provide information
on the energies and nature of the eigenstates not accessible from
linear spectra (see Fig. \ref{fig:Fig1}). However, the results in
Figs. \ref{fig:Fig2} (c) and (d) indicate that more than one vibrational
mode needs to be included explicitly for a quantitative analysis of
the oscillations in FMO.

In this work we have shown that the vibronic exciton model predicts
coherences in FMO with $1.3$ ps dephasing times at $77$ K. Our model
does not invoke static nor dynamic correlations in the site energies
of the pigments, and uses experimentally determined spectral densities
and vibrational frequencies. The long lived coherences are found to
reflect coherent superpositions of vibronic exciton states with dominant
contributions from vibrational excitations on the same pigment. Because
vibrational modes are an inherent property of all pigments, we expect
vibronic excitons to be a general feature in the dynamics of molecular
aggregates. In the exciton language, the long lived coherences reported
here correspond to a coherence between the system and the bath. Because
the resonance coupling in the vibronic exciton model acts on the system
as well as on certain bath modes (Eq. (\ref{eq:HS})), both types
of DOF become mixed. This enhances the effective system-bath coupling
in a way not accounted for in our exciton model. A similar mixing
of system and bath DOF takes place implicitly when the reduced equation
of motion for the electronic system is propagated exactly \cite{Panitchayangkoon2011a}.
It is clear that the mixing takes place for all bath modes. However,
modes of the protein environment are strongly damped and cannot contribute
to a long dephasing time. As shown here, one or more of the underdamped
vibrations found in the BChl monomers are needed to account for the
enhanced dephasing times of the coherences. Our results imply that
the oscillations in the 2D experiment on FMO reflect the dynamics
of the nuclear DOF within a single pigment, which should have little
impact on the transfer of energy from the chlorosome to the reaction
centre. 

\emph{Acknowledgments:} This work was supported by the Wenner-Gren
foundation, \"{O}sterreichischer Austauschdienst (WTZ), Austrian Science
Foundation (FWF), the Swedish Research Council, KAW foundation, Swedish
Energy Agency, the Czech Science Foundation grant GACR 205/10/0989,
and the Ministry of Education, Youth, and Sports of the Czech Republic,
grant MEB 061107. 

\bibliographystyle{prsty}

\begin{thebibliography}{10}

\bibitem{Davydov}
A.~S. Davydov, {\em Biology and Quantum Mechanics} (Pergamon Press, New York,
  1981).

\bibitem{PhotosynthExcitons}
H. van Amerongen, L. Valkunas, and R. van Grondelle, {\em Photosynthetic
  Excitons} (World Scientific, Singapore, 2000).

\bibitem{Kuehn2002a}
O. K\"{u}hn, V. Sundstr\"{o}m, and T. Pullerits, Chem. Phys. {\bf 275},  15
  (2002).

\bibitem{Engel2007a}
G.~S. Engel, T.~R. Calhoun, E.~L. Read, T.-K. Ahn, T. Man\v{c}al, Y.-C. Cheng,
  R.~E. Blankenship, and G.~R. Fleming, Nature {\bf 446},  782  (2007).

\bibitem{Ishizaki2009a}
A. Ishizaki and G.~R. Fleming, Proc. Nat. Acad. Sci. USA {\bf 106},  17255
  (2009).

\bibitem{Hayes2011a}
D. Hayes, J. Wen, G. Panitchayangkoon, R.~E. Blankenship, and G.~S. Engel,
  Faraday Discuss. {\bf 150},  459  (2011).

\bibitem{Calhoun2009a}
T.~R. Calhoun, N.~S. Ginsberg, G.~S. Schlau-Cohen, Y.-C. Cheng, M. Ballottari,
  R. Bassi, and G.~R. Fleming, J. Phys. Chem. B {\bf 113},  16291  (2009).

\bibitem{Collini2009b}
E. Collini, C.~Y. Wong, K.~E. Wilk, P.~M.~G. Curmi, P. Brumer, and G.~D.
  Scholes, Nature {\bf 463},  644  (2010).

\bibitem{Chen2011a}
L. Chen, R. Zheng, Y. Jing, and Q. Shi, J. Chem. Phys. {\bf 134},  194508
  (2011).

\bibitem{Hein2011}
B. Hein, C. Kreisbeck, T. Kramer, and M. Rodriguez, arXiv:1110.1511v2.

\bibitem{Nalbach2011a}
P. Nalbach, D. Braun, and M. Thorwart, Phys. Rev. E {\bf 84},  041926  (2011).

\bibitem{Abramavicius2011a}
D. Abramavicius and S. Mukamel, J. Chem. Phys. {\bf 134},  174504  (2011).

\bibitem{Caycedo2011b}
F. Caycedo-Soler, A.~W. Chin, J. Almeida, S.~F. Huelga, and M.~B. Plenio,
  arXiv:1201.0156v1.

\bibitem{Olbrich2011a}
C. Olbrich, J. Strumpfer, K. Schulten, and U. Kleinekathofer, J. Phys. Chem. B
  {\bf 115},  758  (2011).

\bibitem{Shim2011a}
S. Shim, P. Rebentrost, S. Valleau, and A. Aspuru-Guzik, arXiv:1104.2943v2.

\bibitem{Panitchayangkoon2011a}
G. Panitchayangkoon, D. Voronine, D. Abramavicius, J.~R. Caram, N.~H.~C. Lewis,
  S. Mukamel, and G.~S. Engel, Proc. Nat. Acad. Sci. USA {\bf 108},  20908
  (2011).

\bibitem{MayKuehnBook}
V. May and O. K\"{u}hn, {\em Charge and Energy Transfer Dynamics in Molecular
  Systems} (Wiley-VCH, Berlin, 2001).

\bibitem{Adolphs2006a}
J. Adolphs and T. Renger, Biophys. J. {\bf 91},  2778  (2006).

\bibitem{Christensson2011a}
N. Christensson, F. Milota, J. Hauer, J. Sperling, O. Bixner, A. Nemeth, and H.
  Kauffmann, J. Phys. Chem. B {\bf 115},  5383  (2011).

\bibitem{Wendling2000a}
M. Wendling, T. Pullerits, M.~A. Przyjalgowski, S.~I.~E. Vulto, T.~J. Aartsma,
  R. van Grondelle, and H. van Amerongen, J. Phys. Chem. B {\bf 104},  5825
  (2000).

\bibitem{Polyutov2011a}
S. Polyutov, O. Kuhn, and T. Pullerits, Chem. Phys. {\bf 394},  21  (2012).

\bibitem{Philpott1971a}
M.~R. Philpott, J. Chem. Phys. {\bf 55},  2039  (1971).

\bibitem{Spano2002a}
F.~C. Spano, J. Chem. Phys. {\bf 116},  5877  (2002).

\bibitem{MukamelBook}
S. Mukamel, {\em Principles of nonlinear optical spectroscopy} (Oxford
  University Press, Oxford, 1995).

\bibitem{Zhang1998a}
W.~M. Zhang, T. Meier, V. Chernyak, and S. Mukamel, J. Chem. Phys. {\bf 108},
  7763  (1998).

\bibitem{Tronrud}
D.~E. Tronrud, J. Wen, L. Gay, and R. Blankenship, Photosynth. Res. {\bf 100},
  79  (2009).

\bibitem{Humphrey1996a}
W. Humphrey, A. Dalke, and K. Schulten, J. Mol. Graphics {\bf 14},  33  (1996).

\end{thebibliography}

\end{document}